# Can Blockchain Protect Internet-of-Things?

## A New Concept to Allow Blockchain to Protect Internet-of-Things

Hiroshi Watanabe
Dept. Electrical and Computer Engineering,
National Chiao Tung University
Hsinchu, Taiwan
hwhpnabe@gmail.com

***Abstract*—In the Internet-of-Things, the number of connected devices is expected to be extremely huge, i.e., more than a couple of ten billion. It is however well-known that the security for the Internet-of-Things is still open problem. In particular, it is difficult to certify the identification of connected devices and to prevent the illegal spoofing. It is because the conventional security technologies have advanced for mainly protecting logical network and not for physical network like the Internet-of-Things. In order to protect the Internet-of-Things with advanced security technologies, we propose a new concept (Physical-Logical Link layer) which is a well-designed combination of physical chip identification and blockchain. With a proposed solution of the physical chip identification, the physical addresses of connected devices are uniquely connected to the logical addresses to be protected by blockchain.**

### I. Introduction

Network security has been a hot topic as the computer network becomes spread all over the world. Any existing commercial applications and services would be unseen without modern encryption technology [1]. If data to be exchanged on the network is easily stolen and read by someone, no customers can trust those services. On contrary, if a strong encryption forces customers spend time and money to protect their communication on the network, the services would be detested. Therefore, the security technology has been developed by considering balance between solidity (security) and flexibility (convenience). It is preferable that communication is certainly protected by least effort of customers and lowest cost. Thus, the operation of encryption and protocol may be manipulated by software. However, this simultaneously reduces the cost of hacking. Hackers can connect an attacking target via network without physically moving and then attack the target with software. Once a hacker finds a vulnerability of the protection, the whole hackers all over the world may be able to attack the vulnerable targets. Then, the security system must be revised to resolve the known vulnerabilities. If a hacker finds a new vulnerability, the security system must be revised again. By this way, such revisions are limitlessly repeated.

Following Diffie and Hellman [2], Alice may give her public key to Bob while saving her secret key by herself. Bob can send Alice a letter encrypted by her public key. Only Alice can decrypt and read the letter with her secret key, because no others have her secret key. It is noted here that the letter for Alice must be exposed on the network in order to send it to her. Theoretically, everyone on the network can receive it. However, since none can actually read it other than Alice, we can regard that the letter is delivered to Alice certainly. With this regard, the public key can be turned out to serve as her logical address on the network. As long as the public key encryption is not broken, everyone is able to securely exchange communications on the network at lowest cost.

As illustrated in Fig. 1(a), there are plurality of logical nodes (depicted circles) respectively having logical addresses above logical layer-1. In this logical layer-1, the communication (data transaction) between logical nodes is assumed to be protected by encryption system-1, for example. Suppose that encryption system-1 is broken. The defenders of logical network are forced to fix vulnerability so as to found logical layer-2 to be protected by encryption system-2, as illustrated in Fig. 1(b). If encryption sytem-2 is subsequently broken, logical layer-3 to be protected by encryption system-3 is found on logical layer-2, as illustrated in Fig. 1(c). By this way, logical layers have been repeatedly laminated above datalink layer, every when the security system is reinforced and updated. As an example, the protocols of logical layer-1, -2 and -3 are TCP/IP, https and OAuth, respectively.

A great breakthrough has occurred; which is called blockchain playing a central role of bitcoin technology [3]. The cryptocurrency is thus allowed to be sent directly from one party to another without going through a financial institution. This is Peer-to-Peer (P2P) transaction. Bitcoin is the first realization of this concept. As illustrated in Fig. 2, bitcoin address serves as logical address on the logical network. The transaction among those logical addresses is practically-strictly protected from illegal falsification, as long as blockchain is long enough. Thereby, blockchain has ignited FINTECH 2.0 in most recent years. (Blockchain and its limitation are briefly reviewed in Sec. II.) .

On the other hand, as shown in Fig. 1(a) - (b) and Fig. 2, there are plurality of physical nodes (depicted by squares) respectively having physical addresses below physical layer. The protocol of physical layer may be usually Ethernet as an example. Those physical nodes can exchange data (e.g. frame) with going through physical layer and not through blockchain and not through any logical layers. Therefore, neither

This work is sponsored by IMSolution Co. Ltd.

blockchain nor the conventional security system can protect the data transaction between physical nodes. It is because physical addresses and logical addresses are independent each other.

Indeed, Internet-of-Things (IoT) is a physical network comprising huge number of connected devices (hardware). Those connected devices are physical nodes respectively having physical addresses. In this work, we propose a new concept (Physical-Logical Link) to uniquely link physical addresses to logical addresses with physical chip identification and public key encryption. Thereby, blockchain is allowed to protect IoT. In addition, Physical-Logical Link is fully compatible to blockchain. It means that Physical-Logical Link can be validated anywhere blockchain works.

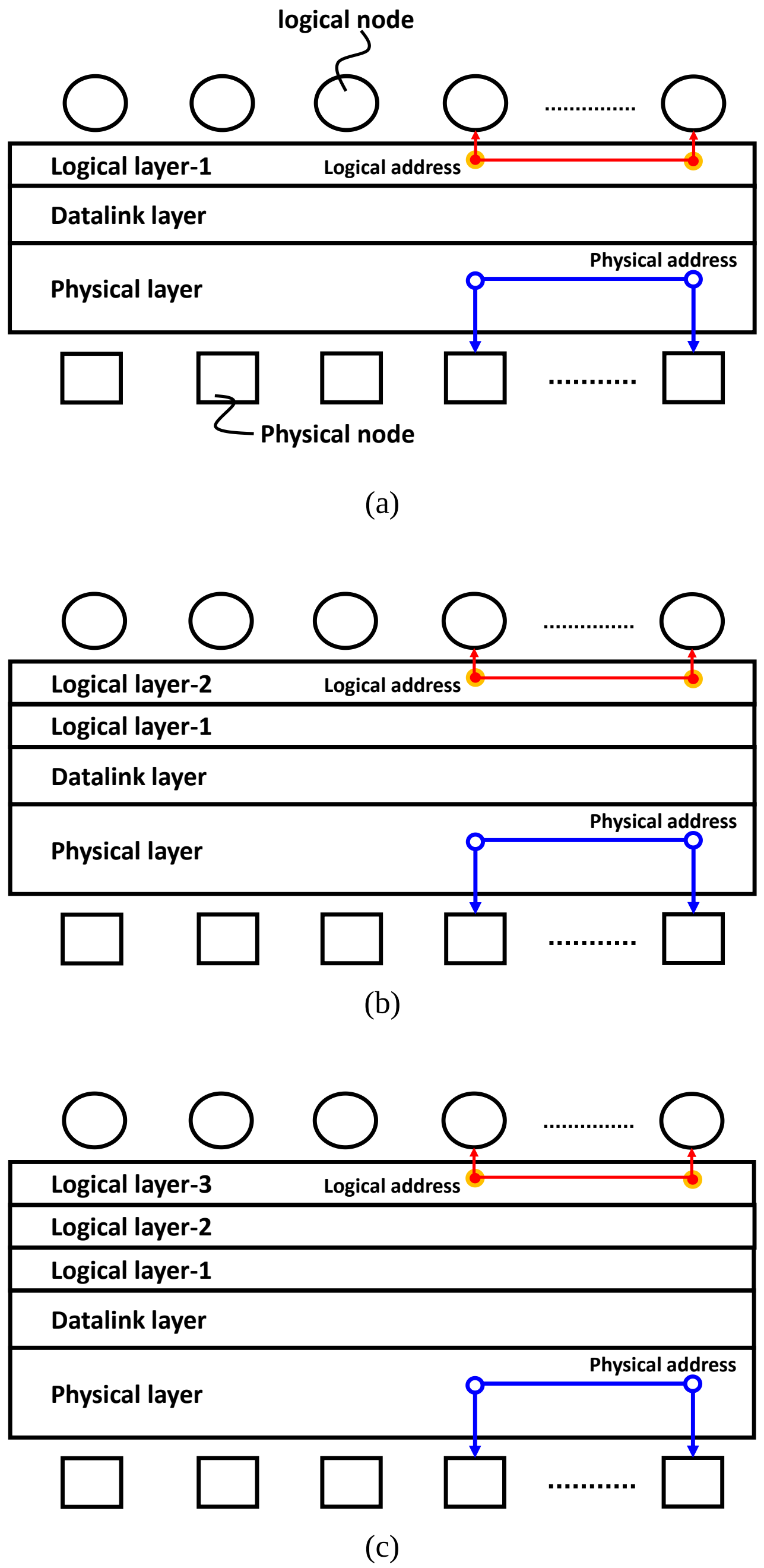

Fig. 1: Communication layers

## II. Brief Review of Blockchain [3]

### A. Transaction

Fig. 3 illustrates a series of transactions from logical node (N-2) to (N-1) and from logical node (N-1) to logical node (N). The bulk squares are transaction units comprising public keys, hash values and electronic signatures. The logical node (N-2) comprises the transaction unit (N-2) and the secret key (N-2); i.e., the public key (N-2), the secret key (N-2), the hash value (N-3), and the electronic signature (N-3). The logical node (N-1) comprises the transaction unit (N-1) and the secret key (N-1); i.e., the public key (N-1), the secret key (N-1), the hash value (N-2), and the electronic signature (N-2). The logical node (N) comprises the transaction unit (N) and the secret key (N); i.e., the public key (N), the secret key (N), the hash value (N-1), and the electronic signature (N-1). The public keys (N-2), (N-1) and (N) make the couples with the secret keys (N-2), (N-1) and (N), respectively. As mentioned above, the public keys (N-2), (N-1) and (N) are the logical addresses (N-2), (N-1) and (N), respectively. The hash value (N-1) is generated from the transaction unit (N-1) by a hash function (e.g., SHA-256). Then, the hash value (N-1) is to be transferred from the logical node (N-1) to (N) with attaching sender's logical address (N-1) and the previous sender's electronic signature (N-2). The hash value (N-2) has been transferred from logical node (N-2) to (N-1) with attaching sender's logical address (N-2) and the previous sender's electronic signature (N-3). The hash value (N-3) has been transferred from logical node (N-3) to (N-2) with attaching sender's logical address (N-3) and the previous sender's electronic signature (N-4), and so forth. At last, the logical address (N) saves the hash value (N-1) with attaching all past senders' logical addresses (N-1), (N-2), (N-3)… and the electronic signatures (N-2), (N-3), (N-4)…, respectively; which is the signed transferring record of the latest data.

The electronic signature (N-1) is generated from the public key (N) and the hash value (N-1) with using the secret key (N-1). It is noted that the electronic signature (N-2) is included into the hash value (N-1) to be sent to the logical node (N). In the logical node (N), we can decrypt the electronic signature (N-1) with using the public key (N-1) and then check if it coincides with a set of the public key (N) and the hash value (N-1). The electronic signature (N-2) is generated from the public key (N-1) and the hash value (N-2) with using secret key (N-2). It is noted that the electronic signature (N-3) is included into the hash value (N-2) to be sent to the logical node (N-1). In the logical node (N-1), we can decrypt the electronic signature (N-2) with using the public key (N-2) and then check if it coincides with a set of the public key (N-1) and the hash value (N-2). And so forth.

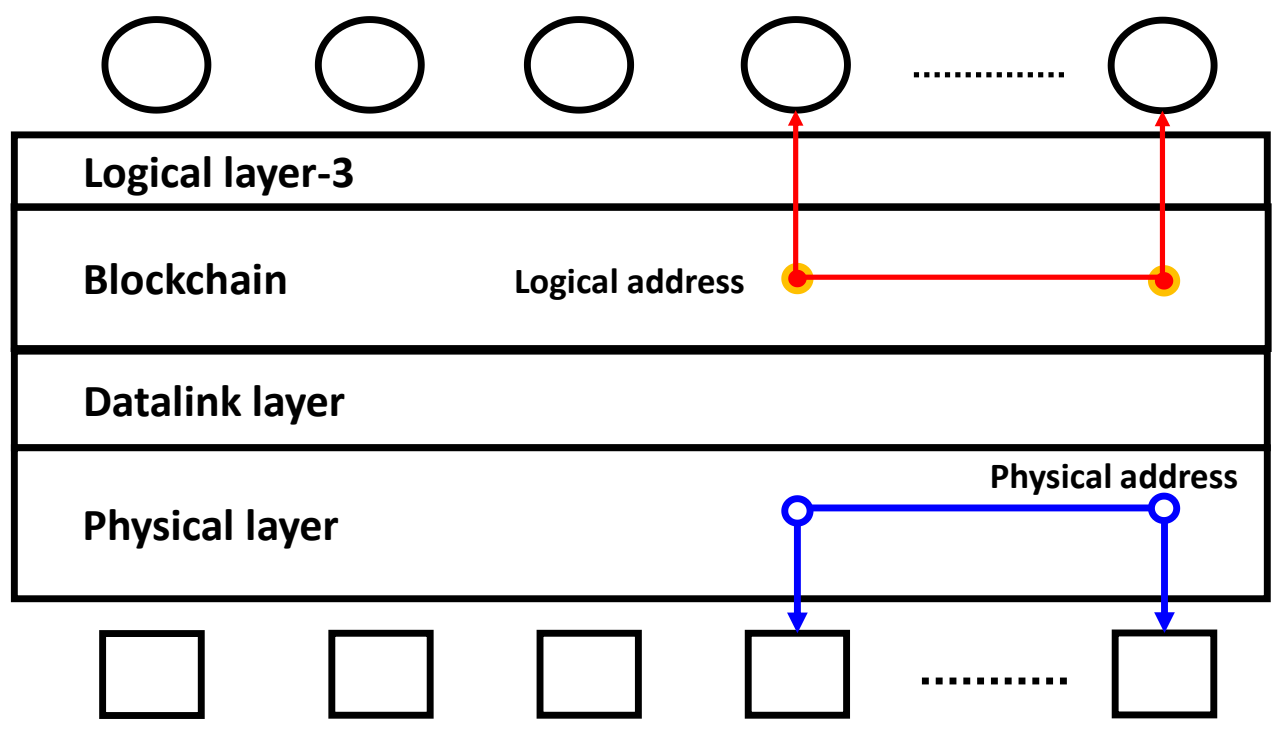

Fig. 2: Communication layer including blockchain

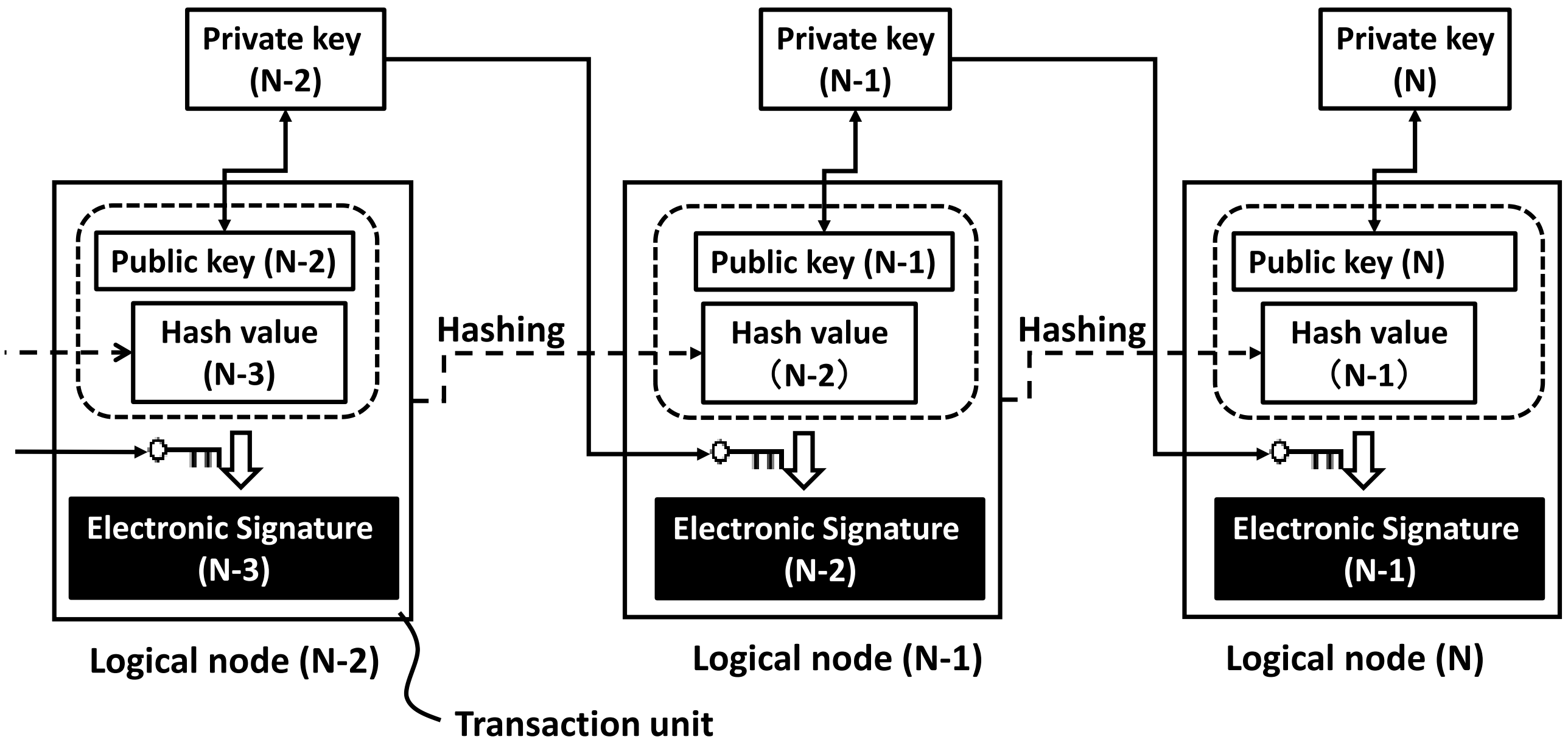

Fig. 3: Data transaction between logical nodes, wherein the dashed arrows denote the hashing.

## *B. Blockchain*

The signed transferring record is thereby included to the latest hash value. In general, a plurality of logical nodes may be able to transfer data to a latest logical node. Then, the trajectory of transferring record may form a tree diagram; which is called Merkle's tree [4], as illustrated in Fig. 4. Each dashed arrow corresponds to the data transfer illustrated in Fig. 3. In this example, there is the latest one at the bottom, which has come from three hash values A, B and C. The hash value A has come from A1 and A2. The hash value A2 has come from A21 and A22. And so forth. No matter how complicated or various the tree diagram is, there is only the latest one at the bottom. This is called "Root of Merkle" and can be the representative of all hash values and all dashed arrows which are included into a same Merkle's tree.

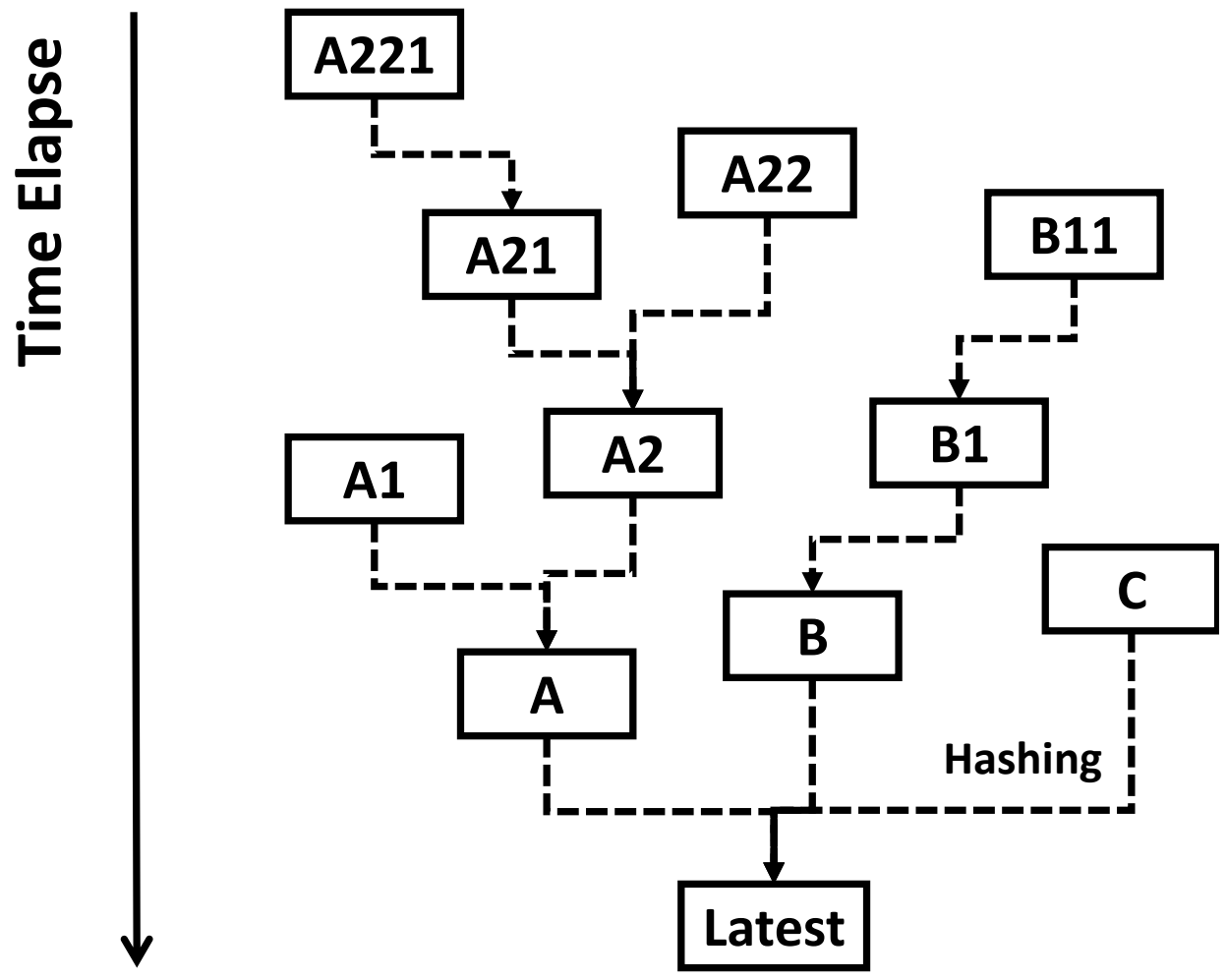

Fig. 4: Merkle's Tree, wherein the dashed arrows denote the hashing related to data transactions. "Latest" at bottom is Root of Merkle.

Usually, several hundred transactions are bunched to form a block labeled by a Root of Merkle. As illustrated in Fig. 5, the block (M-1) comprises the Root of Merkle (M-1), the nonce (M-1), and the block hash (M-2). The block hash (M-1) is generated by hashing the block (M-1), where the nonce (M-1) is tuned to let the first 16 bits be all zero in the block hash (M-1). Since the hash function is irreversible, this tuning costs some computational power. The generated block hash is publicized on the network to be registered. In this example, the block (M-1) is registered by finding the tuned nonce (M-1). The generated block hash (M-1) is publicized on the network and then the Root of Merkle (M-1) is linked to the previous block (M-2) by the hashing. Then, it can be the linkage requirement that the first 16 bits are all zero in the generated block hash. Suppose a user finds a new bunch of transactions having not registered yet, which is labeled by the Root of Merkle (M). If he succeeds to tune nonce (M) to satisfy the linkage requirement, then he can register the new block (M) and then link it to the block (M-1). By repeating this registration of new blocks from the past to the future, the blockchain is formed, as illustrated in Fig. 6.

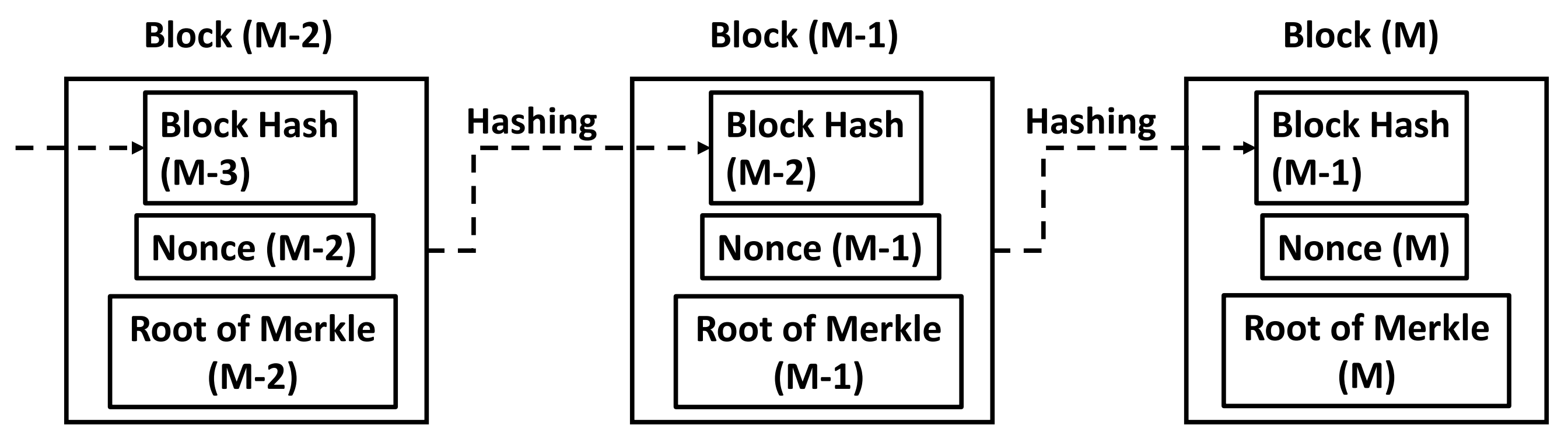

Fig. 5: Linkage of Block-to-Block, wherein the dashed arrows denote the hashing.

### C. *The length of blockchain and the strength of anti-tampering*

Suppose that a hacker tampers a transaction of data related to the hash value A2 in Fig. 4. This tampering influences the hash value A. At last, the latest hash value ("Latest" at bottom) is influenced. This means that tampering a part of transaction record must change the Root of Merkle. Suppose that the Root of Merkle (M-1) is changed by this tampering. This breaks the linkage requirement that the first 16 bits are all zero in the block hash (M-1). In order to recover the linkage, the nonce (M-1) is forced to be revised to keep the linkage requirement from the block (M-1) to (M). However, this must change the block hash (M-1) and then the linkage of blocks (M) and (M+1) is broken. To recover this linkage, the nonce (M) is also forced to be revised. Like this, we are forced to revise nonce (M+1), nonce (M+2), nonce (M+3)… By this way, as the length of blockchain is increased, more computational power must be consumed to tamper a part of transaction record. Therefore, the strength of anti-tampering of transaction record is enhanced as the length of blockchain is increased.

### D. *Limitation of Blockchain Protection*

All logical nodes have logical addresses, respectively. The transaction between those logical nodes are protected by blockchain, as explained in the above. Any applications to be served on the blockchain infrastructure is software. The logical addresses are allocated to accounts of software, respectively. According to the public key encryption, public keys are uniquely coupled to secret keys, respectively, as illustrated in Fig. 7. As long as secret keys are protected, the blockchain can prevent the illegal falsification of data transaction between logical addresses.

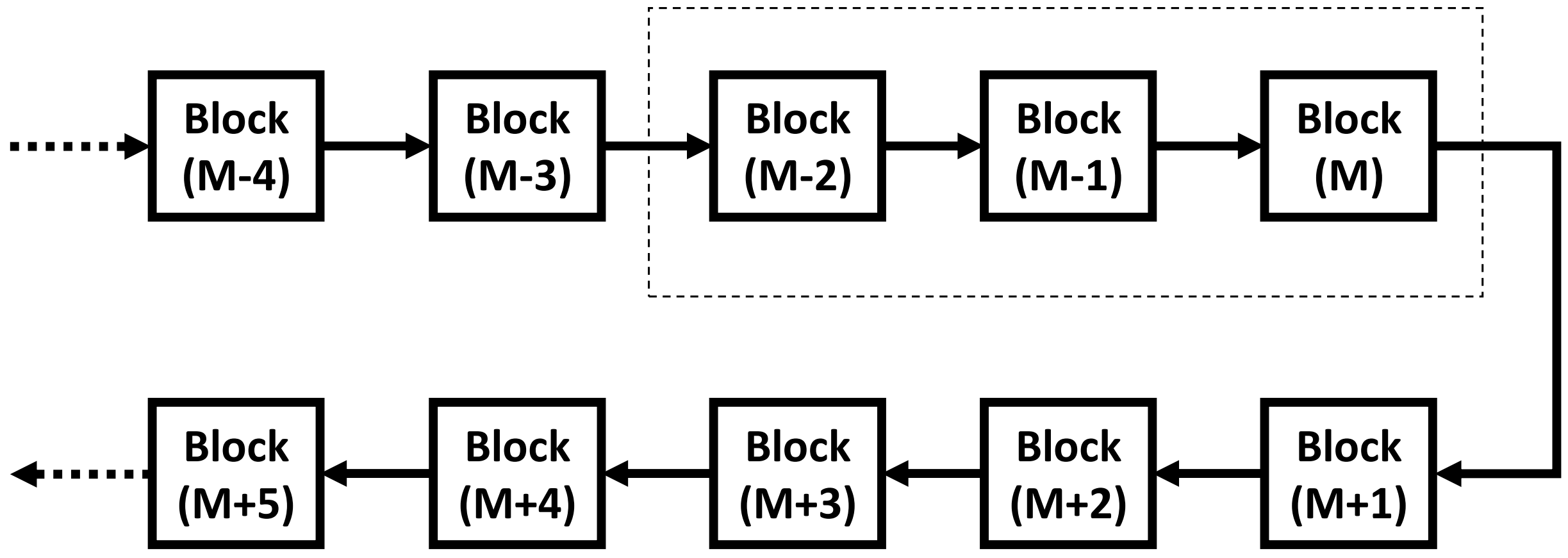

Fig. 6: Blockchain, wherein the dotted square corresponds to the linkage in Fig. 5.

However, the transaction in IoT is performed between physical addresses which are respectively allocated to connected devices (hardware). Therefore, the existing blockchain cannot prevent the illegal falsification of transaction history among connected devices.

## III. NEW CONCEPT TO PROTECT INTERNET-OF-THINGS

As mentioned above, the first serious problem is that physical address is disconnected from logical nodes; then the advanced security technologies protecting those logical nodes cannot protect the data transaction between physical addresses.

In recent years, the majority of connected devices is changing from computers to smaller and various devices; it is expected that there are 5-10 devices per person in average in 2020 [5]. Then, the number of connected devices is drastically increased to become from 26 – 50 billion by 2020 [5], [6]. The serious open issue is therefore how to protect the whole connected devices from the illegal spoofing of physical address.

### A. *Physical Chip Identificaiton*

The physical address having been used extensively is the media-access-controller address (MAC address). However, it is well-known that MAC address is editable; then two different connected devices may be able to have a same MAC address. This means that MAC address is not real physical address. The physical addresses should be therefore uniquely allocated to connected devices, respectively. The data transaction between allocated physical addresses must be always identical to that between those connected devices.

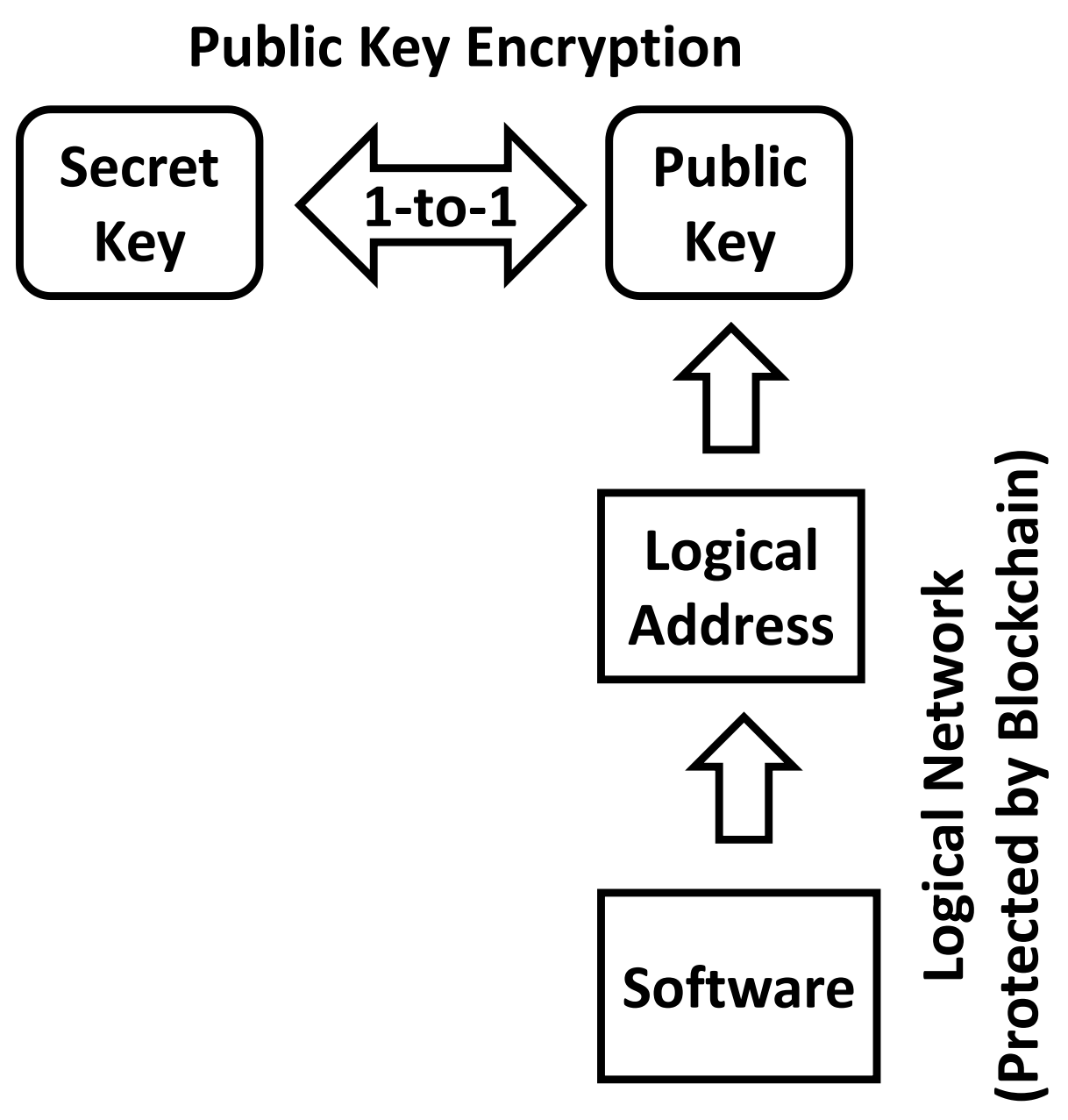

Fig. 7: Data transaction between logical addresses protected by blockchain

```
00000001101001111100000000011010110111000000000111000100000000000000111011011110
0000000100000010100100000000100000101111110000000100011000111100000000100011100100
0000000010001111001110000000100100111010000000001001011011100000000010011001101
0100000001101110010000000000001101111110100000000011100000000110000000111010000
01000000001110100000001100000001000001101000100000001000010100010100000001000010101
0101000000010010011000110000000100100111110010000001001110111010100000001001111111
11110000000101001000000000000001010111001111000000010101111000101000000010110001
101011000000010110011101010000000101110011010100000001100000011001100000001100010
101011100000011010100000000000000011011001001010000000110111001101010000000110111
1010110000000001110011110010100000011110000000101000000011110100000000100000011110
11000110100000001111100100101100000001111101011001100000001111101101110100000001111
11100010100000000001101001111100000000001101011011100000000001110001000000000000001
110110111100000000010000000101001000000001000001011111100000000100011000111100000001
000111001000000000001000111100111000000010010011101000000000100101101110000000000
100110011010100000001101110010000000000001101111110100000000011100000000011000000
01110100000001000000001110100000001100000001000001101000100000001000010100010100000
01000010101010100000001001001100011000000010010011111001000000100111011101010000
00100111111111100000001010010000000000000001010111001111000000010101111000101000
00010110001101011000000010110011101010000000101110011010100000001100000011001100
000011000101010111000000011010100000000000000011011001001010000000110111001101010
000001101111010110000000001110011110010100000011110000000101000000011110100000001
00000011110110001101000000011111001001011000000011111010110011000000011111101101110
1000000111...................
```

Fig. 8: A part of generated physical chip identification code (Chip-ID); which has been turned out to be unchanged even after heating acceleration test at 125 C° for 168 hours. A pair of secret key and public key can be generated from the full version of this code.

A smallest component of connected device is a semiconductor chip. Therefore, to define physical address of a connected device is equivalent to identify a semiconductor chip on the network. It is the physical chip identification. The practical implementation of physical chip identification is to extract physical randomness from mass-product semiconductor memory chip which is included in connected devices. We demonstrate in the conference that the physical chip identification is generated in the manufacturing semiconductor memory chip with no change in front-end process. We call it the identification random access memory (IDRAM) in this report. The IDRAM can generate physical chip identification and also serve as memory chip. Therefore, IDRAM is the cheapest solution for physical chip identification to be easily implemented into connected devices having at least one memory chip. The IDRAM might be similar to physically unclonable function (PUF) [7], [8]. However, the goal of this report is whether or not blockchain can protect IoT. Whether or not IDRAM is a variety of PUF may be discussed elsewhere. Hence, the property of IDRAM is briefly explained below; which has different characteristics from [7], [8].

Fig. 8 is a part of a physical chip identification code (Chip-ID) generated from a mass-product semiconductor memory process. The information quantity of IDRAM is estimated to be about 1.0E1,042,102 if the IDRAM is manufactured with the manufacturing line of a commercial 4Gbit memory products. Even though one-trillion chips are shipped all over the world, the possibility that two different chips accidentally have a same chip identification is about 1.0E–1,042,090. This can be regarded as practically zero. It is also noted that IDRAM can also serve as 4Gbit mass-product memory chip with no practical loss of bit capacity.

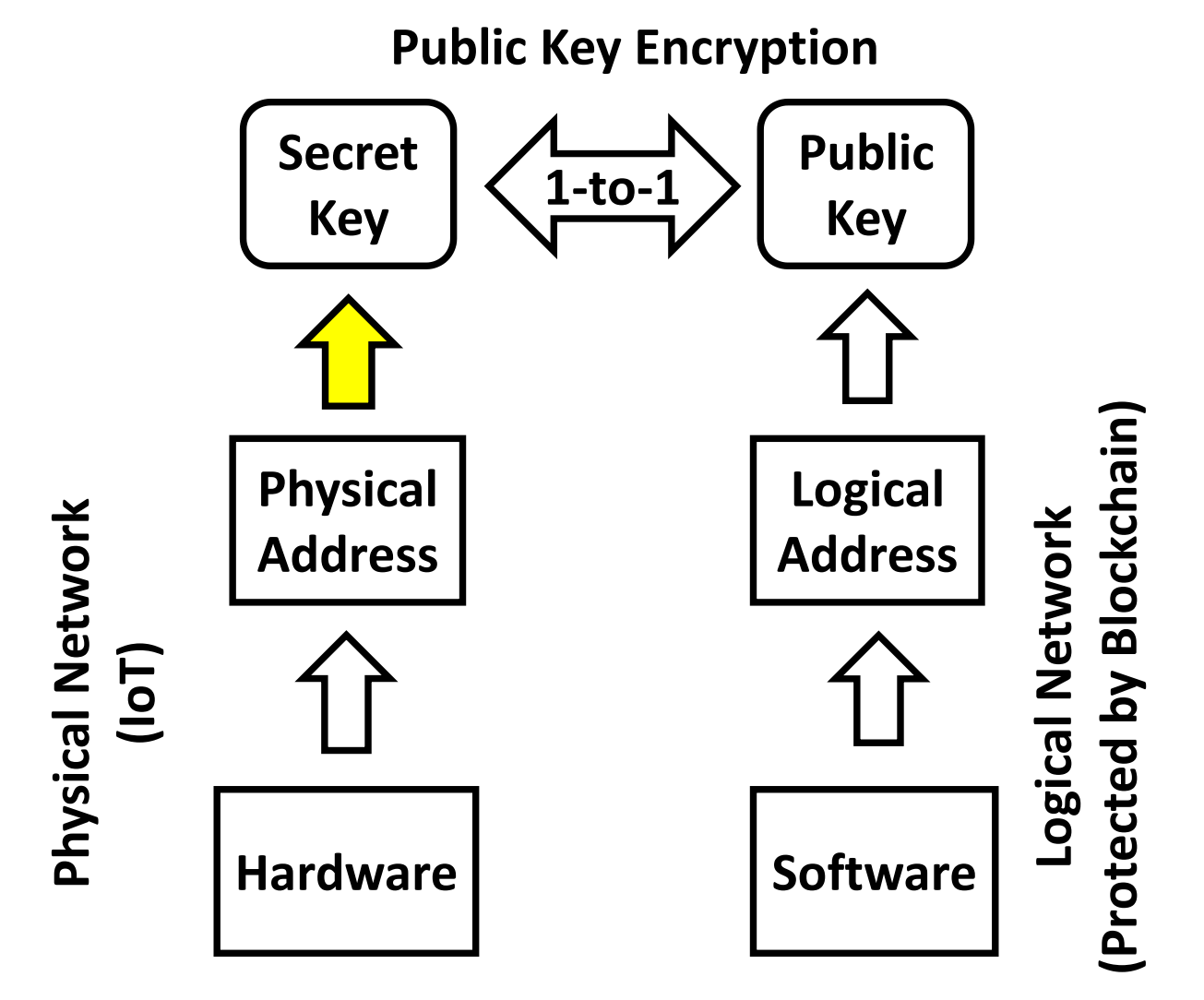

Fig. 9: Linked physical address and logical address

The data retention characteristics is also demonstrated in the conference. First of all, we have already carried out the heating acceleration test of 1,116 IDRAM chips as follows; 1) we measured the identification codes generated from those 1,116 IDRAM chips. Next, these 1,116 IDRAM chips were heated at 125 degree Celsius (C°) for 168 hours. This condition is identical to that for ten years data retention in mass-product flash memories. 3) We measured the identification codes generated from the 1,116 IDRAM chips again. (See Fig. 8. It is a part of the read identification code from one of the 1,116 IDRAM chips.) As a result, it is found that the 1,116 chips respectively generate as the same identification codes as before the heating acceleration. (All 1,116 chips have no inconsistent bits in the identification codes before and after the heating acceleration). This shows *the unprecedentedly good retention characteristics of the physical chip identification.* More detail of IDRAM property will be discusses in the conference.

### B. *Proposed Concept of Physical-Logical Link*

If such a physical address is uniquely linked to a secret key, as illustrated in Fig. 9, it can be uniquely connected to a logical address via the pair of secret key and public key. Thereby, to protect the data transaction between logical addresses becomes identical to protect that between physical addresses. In other words, the data transaction between physical addresses may be able to be protected by blockchain.

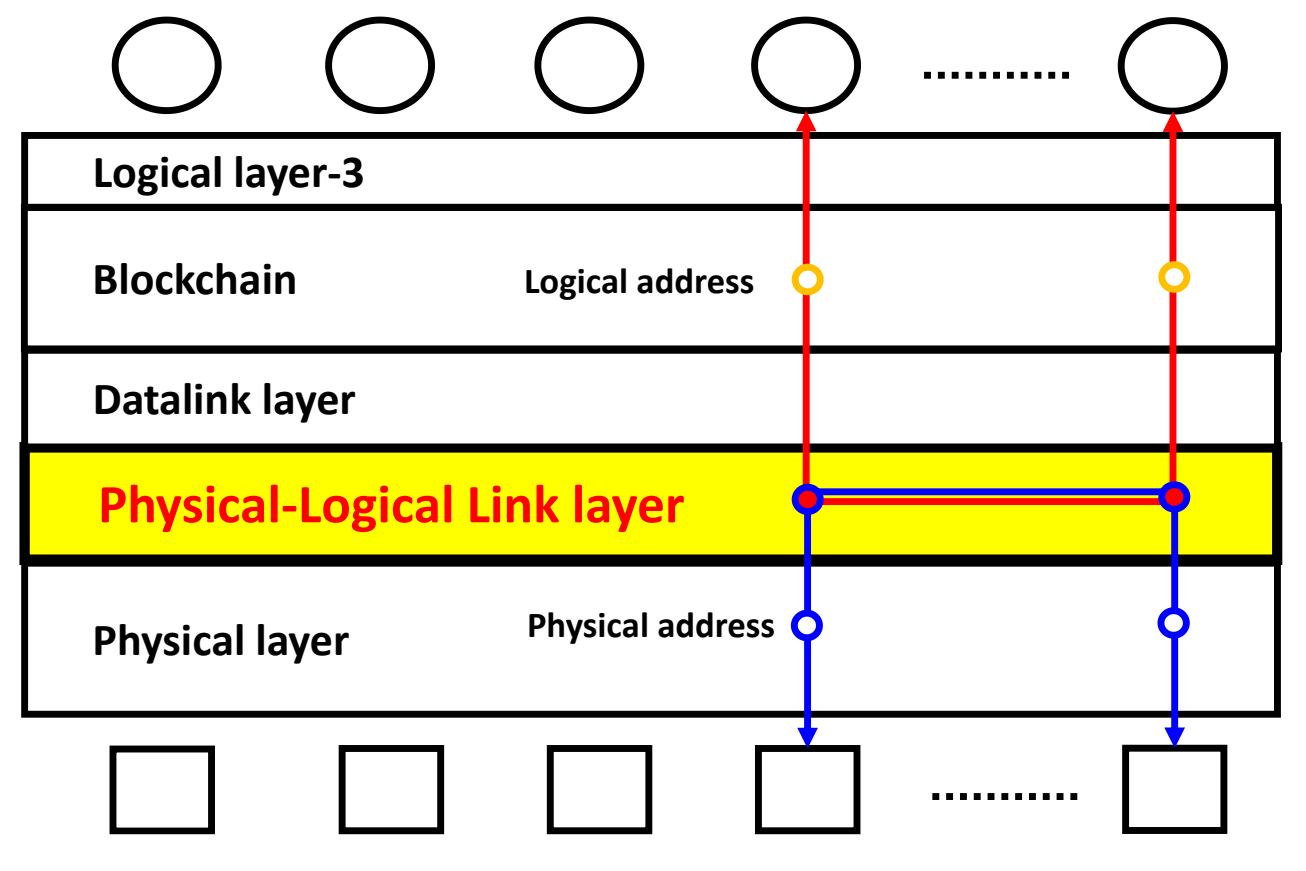

Fig. 10: Communication layer including Physical-Logical Link layer

The physical chip identification is installed by replacing memory chip included in connected device with IDRAM chip. It is also noted that no change in application software is necessary to this installation. Since upper layer is free from any change in lower layers in the communication layer structure, a new layer can be inserted into below datalink layer, as illustrated in Fig. 10. We may call it Physical-Logical Link layer in this report.

By this way, the data transaction between connected devices is carried out with going through Physical-Logical Link layer. Additionally, the logical addresses are respectively combined with the physical addresses in Physical-Logical Link layer. As long as the data transaction between those logical addresses is protected by blockchain, the data transaction between those physical addresses can be protected by blockchain. This is the concept of Physical-Logical Link layer.

Fig. 11 illustrates the data transaction with going through Physical-Logical Link layer more in detail. The upper squares are the identification cores (ID cores) and the lower ones are transaction units which are identical to those in Fig. 3, respectively. The ID core comprises a secret key, a Chip-ID and a key generator. The Chip-ID is some kind of code to carry out the physical chip identification, i.e., to define physical address of a semiconductor chip. More concretely, it is a code to be generated from physical randomness which is characteristic to a semiconductor chip. This Chip-ID is input to the key generator and the key generator generates a set of public key and secret key which are uniquely coupled according to the algorithm of Rivest, Shamir and Adelman (RSA) [9], for example.

In the RSA method, a non-zero positive integer $e$ may be given at first. This $e$ is usually $1 + 2E16 = 65537$. Next, a set of large prime numbers $(p, q)$ are generated in some kind of method. Then, the product of those prime numbers, i.e., $n = pq$, is calculated. Hence, $(e, n)$ is the public key. Subsequently, we look for an integer $\phi$, where the reminder of dividing $\phi$ by $(p-1)(q-1)$ is $1$. At last, the secret key $d$ is obtained by dividing this $\phi$ by $e$. It is noted that this $(p, q)$ must be confidential.

We may generate a set of large prime numbers $(p, q)$ from physical randomness of semiconductor chip, and then delete it after generating the secret key. For example, we may add an arbitrarily selected positive integer to Chip-ID, and then check if the sum is a prime number or not. If it is a prime number, we may define it as $p$. Next, we may add another arbitrarily selected positive integer to Chip-ID and then check if the sum is a prime number or not. If it is a prime number, we may define it as $q$. It is also necessary to check if $p$ and $q$ are unequal. Those positive integers to be added to Chip-ID may be given by random-number generator, given by an external input, or confidentially stored ahead in semiconductor chip. However, it is preferable that those positive integers are prime numbers each other.

Another algorithm to generate secret key from Chip-ID is Elgamal's method [10]. At first, a large prime number $p$ and its primitive root $g$ are given. Next, a hashed chip-ID may be divided by $p-1$ to form a secret key $x$ with some arithmetic manipulation. Subsequently, the primitive root $g$ to the power of $x$ may be divided by the prime number $p$. The reminder of this division may be a public key. Since the secret key is generated before generating the public key, Fig. 11 may be replaced by Fig. 12.

As long as Chip-ID linked to a secret key can be regarded as physical address of the connected device, the physical address is successfully linked to the logical address, as illustrated in Fig. 9.

In addition, if removing Chip-IDs and key generators from ID cores, Fig. 11 and Fig. 12 are identical to Fig. 3. It is noted that Chip-IDs and key generators are respectively included in semiconductor chips. By this way, we can also find that the installation of Physical-Logical Link layer is fully compatible to that with going through blockchain.

### C. Appplications

All connected devices having a memory chip is possible to implement this IDRAM. Such a memory chip may be a stand-alone chip of, or, an embedded chip including the followings: DRAM, MRAM, ReRAM, PCRAM, FRAM, NOR Flash, NAND Flash, Mask ROM, Junction ROM and so forth. Those are either stand-alone or embedded chips of volatile memories to non-volatile memories. Among them, the DRAM is a cheapest solution to implement the IDRAM [11]. The DRAM has been and will be extensively used to serve as main memories for any kind of processor units (CPU, MPU, GPU, GPGPU…) and cash memories for any kind of controllers. Accordingly, the stable shipment in large quantity can be expected. This is decisively preferable to cover as the majority of IoT as possible. For one of examples, the application to SSD controllers is presented in the Flash Memory Summit 2017 [12]. The cash memories (DRAM chips) of SSD controllers are replaced by IDRAMs with no change of the front-end-process in the manufacturing of DRAM; such that physical addresses and logical addresses are respectively connected in the Physical-Logical Link layer, as illustrated in Fig. 10. This certainly prohibits the spoofing of physical addresses. Thereby, blockchain is allowed to protect data transaction among connected devices having SSDs. In a similar way, blockchain is allowed to protect data transaction among connected devices respectively having any kind of memory chips with IDRAMs, as mentioned above. For example, the data transaction between a connected device with DRAM and another connected device with MRAM is protected. The PUF is not always necessary, if blockchain can successfully cooperate with physical chip-IDs.

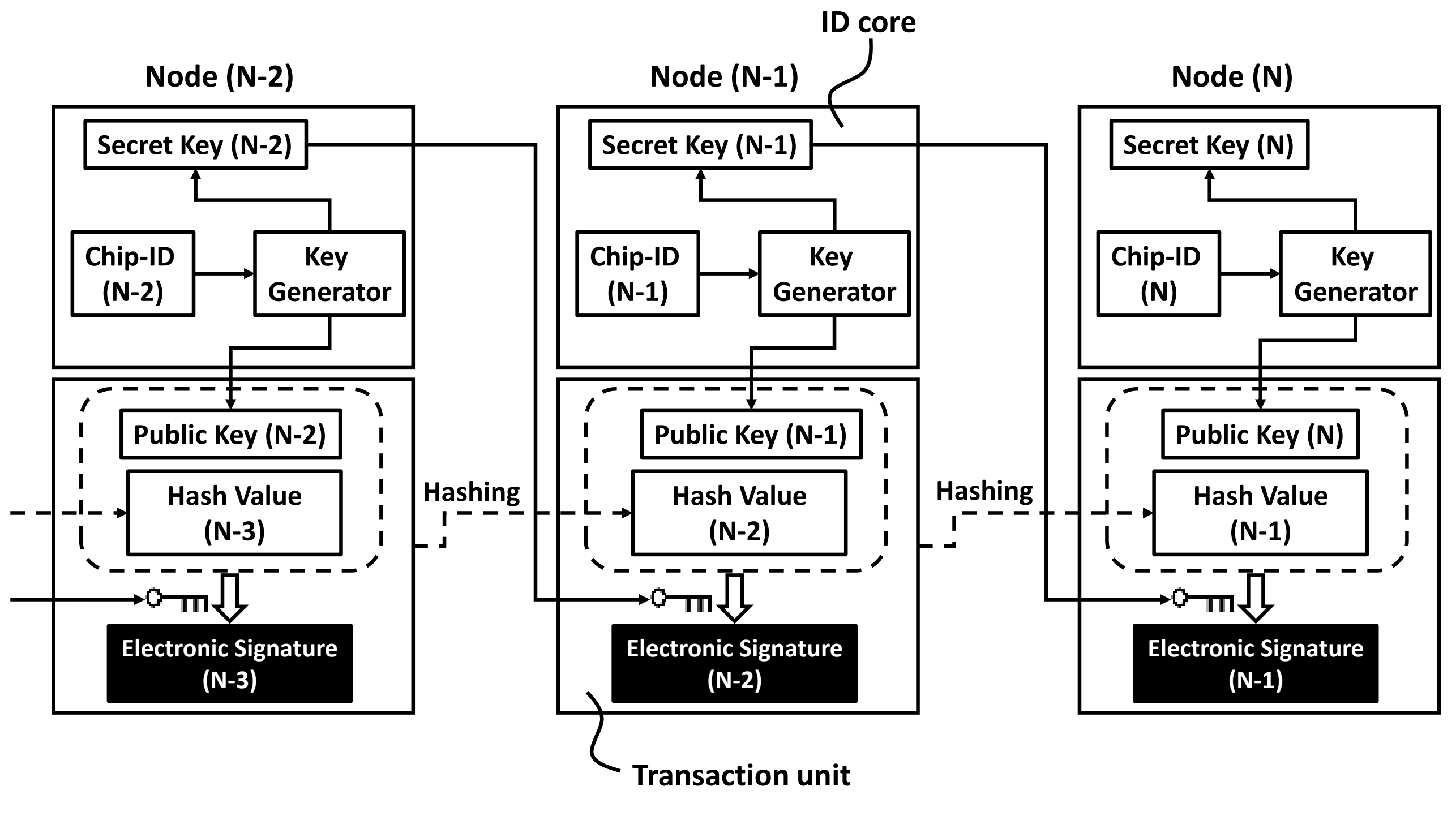

Fig. 11: Data transaction with going through Physical-Logical Link layer, wherein the dashed arrow denote the hashing.

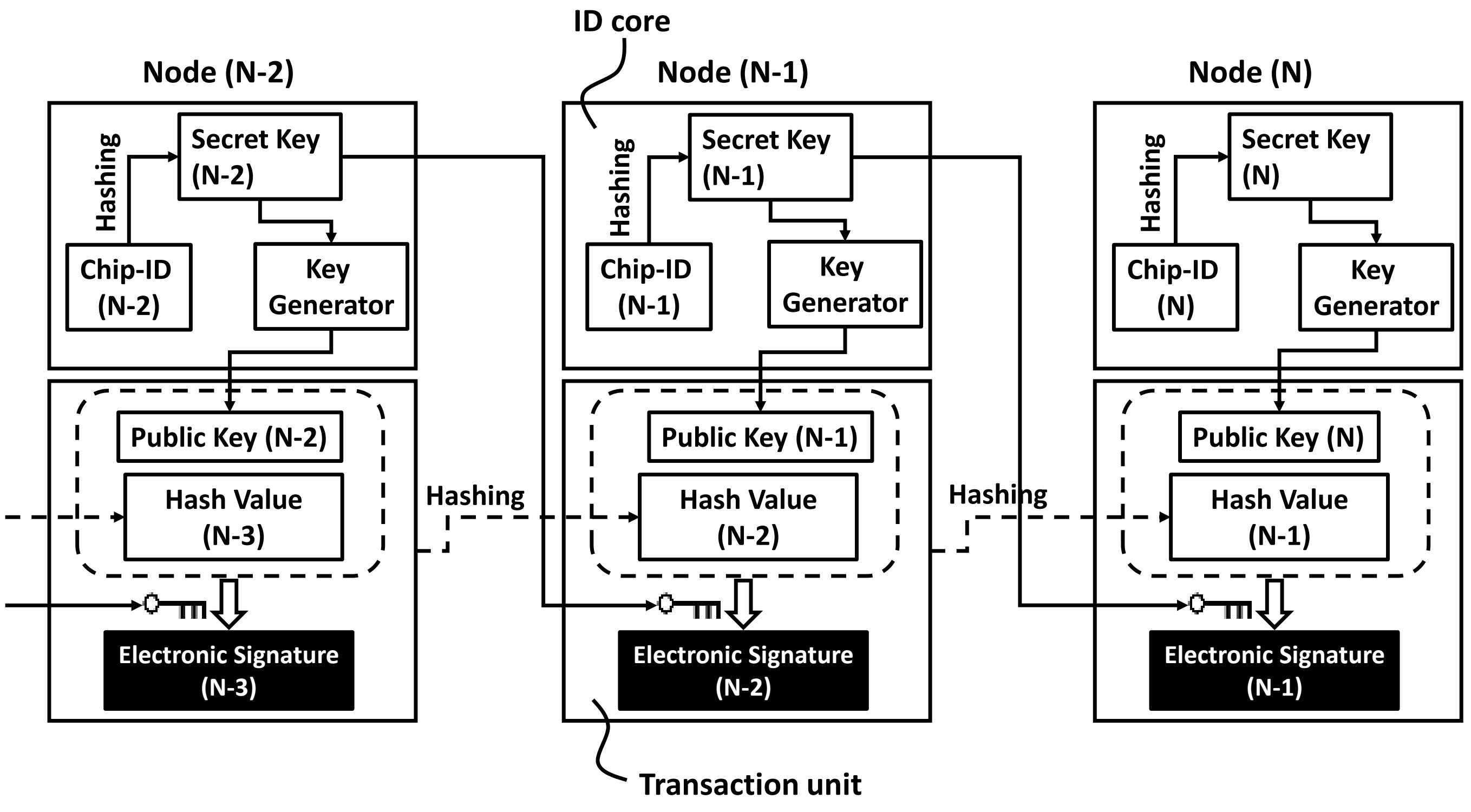

Fig. 12: Data transaction with going through Physical-Logical Link layer, wherein the dashed arrow denotes the hashing.

## IV. Conclusions

IoT is a physical network. The physical network comprises huge number of connected devices, i.e., physical nodes respectively having physical addresses. However, it is still an open problem to protect data transaction between physical nodes from the illegal spoofing. On the other hand, the logical network comprises huge number of logical nodes respectively having logical addresses. The data transaction between logical addresses can be protected by blockchain practically-certainly. We then propose IDRAM; which can uniquely link physical address to logical address. This allows blockchain to protect the data transaction between physical nodes from the spoofing, i.e., to protect IoT.

The IDRAM is a chip to generate physical address from physical randomness related to the manufacturing of memory chip. The installation of IDRAM is very easy. We may replace memory chip included in connected device with IDRAM chip, while no change is required in the application software working thereon. This means that the installation of IDRAM is fully compatible to blockchain as well as it is the installation of Physical-Logical Link layer. The possibility that two different IDRAM chips accidentally have a same physical address is about 1.0E−1,042,090 (practically zero), as long as the IDRAM chips are manufactured in the process line of a 4Gbit commercial memory products. Additionally, the IDRAM chip can be manufactured with no change in the front-end-process of the mass-product memory chip.

## Acknowledgment

The author would like to express his sincere appreciation to Y. Nagai, A. Tseng, R. Shirota, T. Hamamoto, and A. Kinoshita for their fruitful suggestions and helpful discussions.